\documentclass[twocolumn,showpacs,preprintnumbers,amsmath,amssymb]{revtex4}
\usepackage{graphicx}

\begin{document}
\title{Spin Model for Inverse Melting and Inverse Glass Transition}
\author{Nurith Schupper and Nadav M. Shnerb}
\affiliation{Department of Physics, Bar-Ilan University, Ramat-Gan
52900 Israel}
\begin{abstract}
A spin model that displays inverse melting and inverse glass
transition is presented and analyzed. Strong degeneracy of the
interacting states of an individual spin leads to entropic
preference of the "ferromagnetic" phase, while lower energy
associated with the non-interacting states yields a "paramagnetic"
phase as temperature decreases. An infinite range model is solved
analytically for constant paramagnetic exchange interaction, while
for its random exchange, analogous results based on the  replica
symmetric solution are presented.  The qualitative features of
this model are shown to resemble a large class of inverse melting
phenomena. First and second order transition regimes are
identified.
\end{abstract}


\maketitle

We all tend to associate order parameter with order, namely, with
less entropic microscopic realizations. This is indeed the general
situation in nature: crystals are more  ordered  than liquids,
ferromagnets have less entropy than paramagnets. Even the entropy
associated with a glass, an out of equilibrium, frozen frustrated
state, is less than that of a liquid phase of the same material.

There are, however, exceptions, where an "order parameter" does
not reflect order, and the entropy growth during crystallization
or freezing. The prototype of these phenomena is \emph{inverse
melting}, i.e., a reversible transition between a liquid phase at
low temperatures to a high temperature  crystalline phase,
observed in $He^3$ and $He^4$ at extreme conditions (temperature
below $1^\circ$K, pressure above 25 bar) \cite{helium}. A similar
phenomenon was observed recently at room temperature and
atmospheric pressure in P4MP1 polymer solutions \cite{greer}.
Ferroelectricity in Rochelle salt is another example, where the
spontaneous polarization is lost below the (lower) Curie
temperature, this time the transition is second order in type
\cite{rochelle}. The pinned-crystalline inverse transition of
vortex lines in the presence of point disorder at high temperature
superconductors \cite{pinning} is also considered as an example of
inverse melting. However, in that system, the intensive order
parameter (bulk magnetization) is lower in the crystalline phase,
and the response functions are higher, i.e., the disordered phase
is stiffer than the ordered phase.

Even if the crystalline state is the thermodynamically preferred
one, the dynamics of the system may prevent its appearance. In
glass forming materials ergodicity breaking takes place at a
finite temperature and the system is trapped into a frozen
disordered state. One  expects that an "inverse" glass transition
phenomenon, analogous to inverse melting, may also take place. An
interesting example in polymeric systems is the reversible
thermogelation of Methyl Cellulose solution in water
\cite{Chevillard}. When a (soft and transparent) solution of
Methyl Cellulose is heated (above $50^\circ$C, for a 10 gr/liter
solution) it turns into a white, turbid and mechanically strong
gel. Unlike the boiling of an egg that involves an irreversible
transition from a metastable to a stable state, this transition is
reversible upon cooling, and the polymer is  redissolved on
subsequent cooling. In its high temperature phase, the Methyl
Cellulose gel exhibits, like many other gels \cite{gel}, glassy
features.  Non monotonic temperature dependence of the glassy
order parameter has been already reported for a random
heteropolymer in a disordered medium \cite{randompoly}; this may
be considered as the glassy analogue of the flux line
crystallization \cite{pinning}. The liquid-liquid transition
theory for polyamorphous materials predicts an inverse freezing
transition even for the most known liquid, water. In the
hypothesized phase diagram presented in \cite{stanley} a low
density liquid (at about 150 bar, $-100^\circ$C) becomes a low
density amorphous ice upon heating.

In many branches of statistical physics the presentation of a
simple spin model (Ising, Potts, and SK models, for example) turns
out to be a very beneficial step that yields both physical insight
and quantitative predictions. In this paper, such a model for
inverse melting is  presented and analyzed for homogenous and
heterogenous systems in the mean field level. The model exhibits
both inverse melting and inverse glass transition, and allows
first order and second order transitions. We believe that this
generic model is applicable for the qualitative description of the
above mentioned phase transitions (except for the inverse melting
in superconductors which requires a different model).

\begin{figure}
  \includegraphics[width=7.7cm]{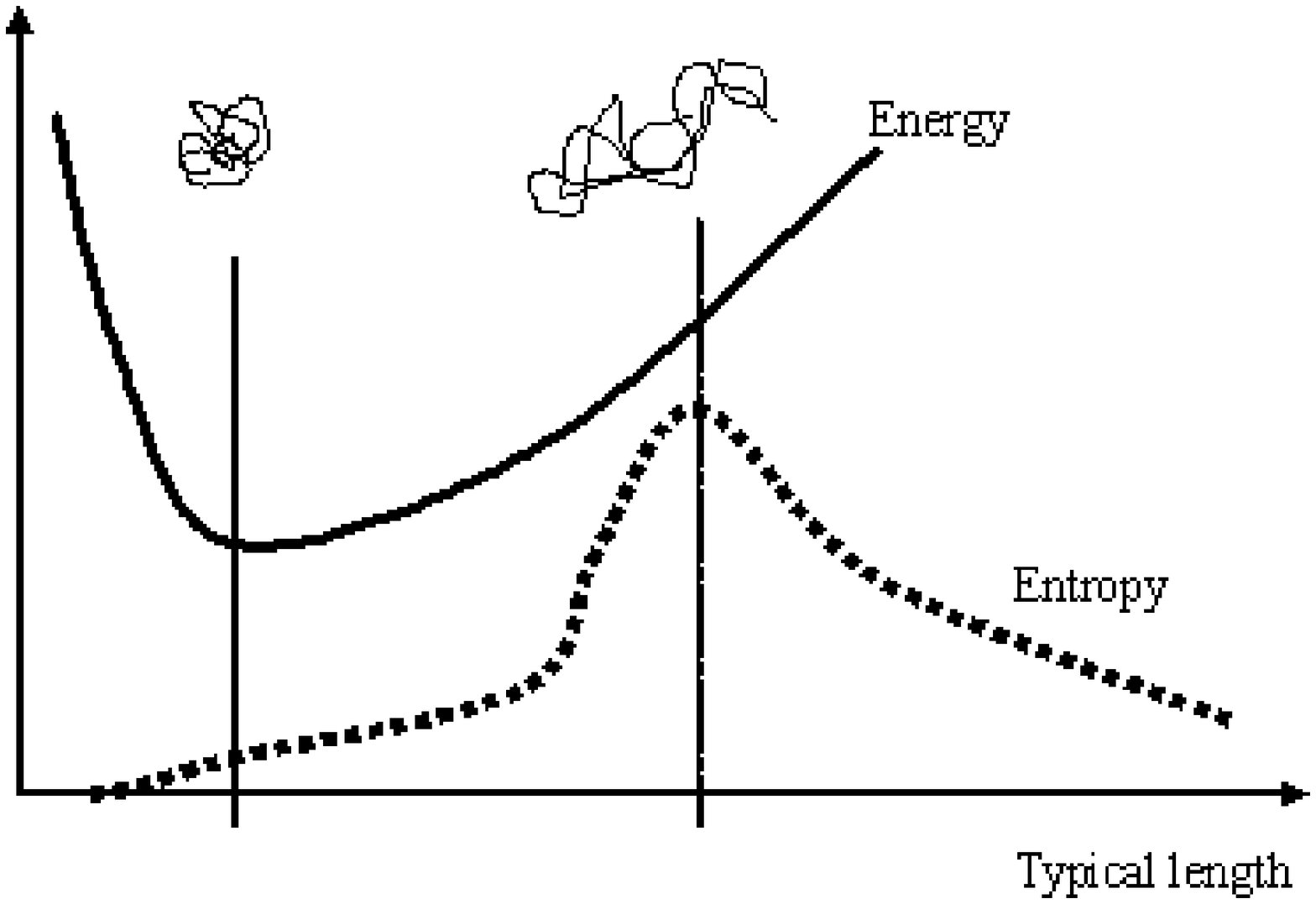}\\
  \caption
{A sketch  of the energy and entropy dependence on the linear size
of a Methyl Cellulose polymer in water. The folded conformation
costs less energy due to more favorable interactions between
hydrophobic sequences along a single chain, but are less entropic
as water molecules has to arrange in cage like structures around
the hydrophobic constituents of the chain. The unfolded
conformation admits much more microscopic configurations. The
interaction with other polymers in the solution is suppressed in
the folded state.} \label{fig1}
\end{figure}

Let us begin with an  intuitive argument focusing on one of the
above mentioned systems, namely,  a single Methyl Cellulose
polymer  chain in water. In order to explain the inverse freezing
it seems plausible to assume that its folded conformation is
favored energetically while its unfolded conformation is favored
entropicaly [See figure (\ref{fig1})]. The entropy growth of the
open conformation may be related to the number of possible
microscopic configurations of the polymer itself, but it may be
attributed also to the spatial arrangement of the water molecules
in its vicinity \cite{haque}.

The main cause for inverse freezing is that the "open"
conformations of the polymer are also the \emph{interacting}
structures, as they allow for the formation of hydrophobic links
with other polymers in the solution, a process that leads to
gelation. This seems to be a general prescription to both inverse
melting and inverse glass transitions: the noninteracting state is
favored energetically, while the interacting state is favored by
the entropy.

Let us now present a very simple model that incorporates these
features. It is based on the Blume-Capel model
\cite{Blume},\cite{Capel} for a spin one particle with "lattice
field" that lower the energy of the "zero" (noninteracting) state.
In contrast with the original Blume-Capel model, we consider the
$\pm 1$ spin states (that interact with other spins) to be more
degenerate.  The system consists of a lattice of N sites and the
Hamiltonian is given by
\begin{eqnarray}
\label{eq:1} H=-J\sum_{<i,j>} S_{i}S_{j}+D\sum_{i=1}^N S_{i}^2
\end{eqnarray}
where the spin variables are allowed to assume the values
$S_i=0,\pm 1$. The summation over $<i,j>$ is over each distinct
pair once. Turning back to our polymer analogy, spin $0$
represents schematically  the compact non-interacting polymer
coil, the stretched polymer (interacting with its neighbors) is
represented by spin $\pm 1$. The positive constant $D$  measures
the energy preference of the compact spatial configurations, and
the "ferromagnetic" interaction between spins, $J$, is related to
the concentration of polymers (or the pressure). The $0$ spin
state is assumed to be  n-fold degenerate, and the $\pm 1$ states
are m-fold degenerate so that $r=m/n \geq 1$ is the degeneracy
ratio that dictates the entropic advantage of the interacting
states. It turns out that all the results presented here are
independent of the absolute degeneracies $m$ and $n$, and depend
only on their ratio $r$.

\begin{figure}
  \includegraphics[width=7.7cm]{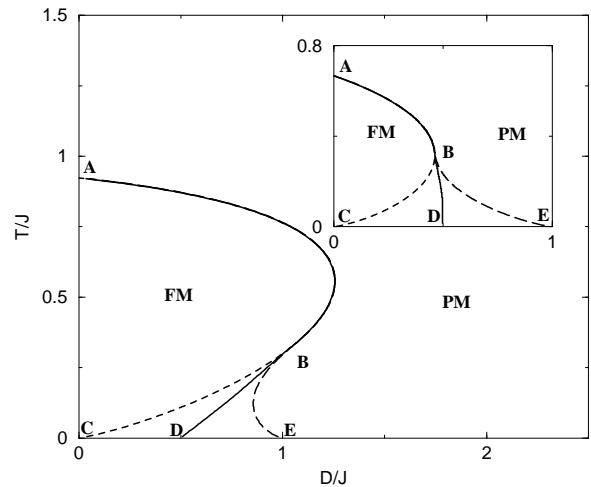}\\
  \caption{Phase diagram and spinodal lines for the ordered
  model Eq. (\ref{diff}) in the $D-T$ plane
  for $r=1$  (Blume-Capel model, inset) and for $r=6$.
  The value of $r=6$ has been chosen in order for the effect to be more pronounced,
  but inverse melting  is seen for $r$ lower than $2$ } \label{fig3}
\end{figure}

Using standard gaussian integral techniques one finds an
expression for the free energy per spin in the infinite range
limit:
\begin{eqnarray} \label{diff}
\label{eq:2} f \equiv F/N  =\beta J m^2/2-ln[1+2 \  r \ cosh(\beta
J m)e^{-\beta D}]
\end{eqnarray}
\noindent where m is the order parameter of the system
(magnetization per spin), $m \equiv \langle \frac{1}{N}
\sum_{i=1}^N S_i \rangle$. The phase transition curves are
obtained numerically by solving for the minimum of Eq.
(\ref{diff}) with respect to $m$. Scaling the temperature and $D$
with the interaction strength $J$, the phase diagram is shown in
Figure (\ref{fig3}). In the inset, results are presented for the
original Blume-Capel model (i.e., the $r=1$ case): the line AB is
a second order transition line, above it is a paramagnetic ($m=0$)
phase and below it the system is  Ferromagnetic ($m \neq 0$).
Below the tricritical point (B) the phase transition is first
order, and the three lines plotted are: the spinodal line of the
ferromagnetic phase BE (above this line the $m\neq0$ solution
ceases to exist), the spinodal line of the paramagnetic phase BC
(below this line there is no $m=0$ minimum of the free energy) and
the first order transition line BD. Along BD  the free energy of
the paramagnetic phase is equal to that of the ferromagnetic
state. Clearly, the Blume-Capel model displays  no inverse
melting: an increase of the temperature induces smaller order
parameter.

The situation is different as $r$ increases, as emphasized by the
main part  of Figure (\ref{fig3}). The same phase diagram is
presented, but now $r=6$, so the interacting states have larger
entropy. The tricritical point is shifted to the left, leaving a
region of second order inverse melting, and the orientation of the
BD line also changes, establishing the possibility of first order
inverse melting. Note that the $r=6$ transition  lines converge to
the $r=1$ lines as  $T \to 0$, since  the entropy has no effect on
the free energy at that limit. The ferromagnetic phase also covers
larger area of the phase diagram for $r=6$, a fact that reflects
again its entropic advantage.

\begin{figure}
  \includegraphics[width=7.7cm]{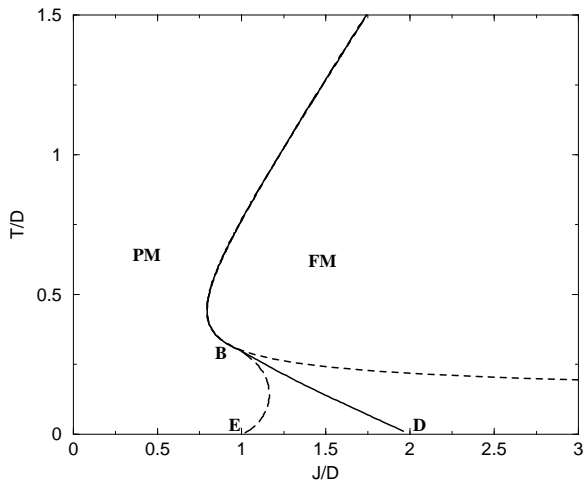}\\
  \caption
 {Phase diagram and the spinodal lines for
  the ordered Blume-Capel model in the  interaction-temperature plane
  with $r=6$. The interaction J/D represents the concentration ("pressure").}
  \label{fig4}
\end{figure}

To allow qualitative comparison of our cartoon model with
experimental results, the appropriate parameters should be
identified. There are three parameters in the model as it stands:
$D$ represents the energetic advantage of the noninteracting
state, $r$ (if larger than 1) is the entropic gain of the
interacting state, and $J$ is the strength of the interaction. In
most of the physical systems that display inverse melting the
controlled external parameter is the strength of the interaction:
pressure (for $He^3$ and $He^4$) or concentration of the
interacting objects (for polymeric systems and Rochelle salt -
Ammonium Rochelle salt mixtures). As long as the only effect of
the pressure is to increase the strength of the effective
interaction among constituents, it may be modelled by changing
$J$. The resulting phase diagram should be compared, though,  with
the $T-J$ plot of our model presented in Figure (\ref{fig4}). The
decrease of the transition temperature with the increase of
interaction strength (pressure) is physically intuitive, as larger
interaction favors energetically the ferromagnetic phase. As
emphasized recently by \cite{stillinger}, the slope of the first
order transition line in the pressure-temperature plane is
required by the corresponding Clausius-Clapeyron equation:
\begin{eqnarray}
\label{eq:3}  \frac{d P}{d T}=\frac{S_2-S_1}{V_2-V_1}
\end{eqnarray}
where $V_2$, $V_1$ are the volume (the extensive parameter
conjugate to the pressure) of the solid and liquid phases
respectively, and $S_2$, $S_1$ are their entropies. Inverse
melting is possible if the numerator of (\ref{eq:3}) is negative,
so for "normal" transitions ($V_2
> V_1$) one expects a negative slope of the transition line.
In real magnetic or electric system the intensive-extensive pairs
[magnetization-magnetic field ($\textbf{M} \cdot d\textbf{H}$) or
polarization-electric field ($\textbf{P}\cdot d\textbf{E}$)],
appear in the free energy function with inverse sign relative to
$PdV$. If the order parameter vanishes, or takes smaller values,
in the "liquid" (disordered) phase, this implies also negative
slope of the first order transition line in the
temperature-external field plane. An interesting exception is the
inverse melting of vortex liquid in superconductors, where the
magnetization of the crystalline phase is smaller than that of the
liquid and the transition line slope is actually positive.

Inverse freezing, the (reversible) appearance of glassy features
in a system upon raising the temperature, may be incorporated in
our model by  introducing random coupling $J_{ij}$, as in the
standard spin-glass models \cite{binder}. This randomness may fit,
in particular, to the gelation transition of Methyl Cellulose, as
it occur \emph{only} when the hydrophobic sequences  are deposited
at random along the chain. The random-exchange analogue of the
Hamiltonian (\ref{eq:1}) is:
\begin{eqnarray}
\label{eq:4} H=\sum_{<i,j>} J_{ij}S_{i}S_{j}+\sum_{i=1}^N D
S_{i}^2
\end{eqnarray}
where the exchange interaction between the $i$ and the $j$ spin is
taken at random from some predetermined distribution. Following
the paradigmatic Sherrington-Kirkpatrick (SK) analysis
\cite{binder} of the infinite range  spin glass, we assume
gaussian distribution of the exchange term with zero mean:
\begin{eqnarray}
\label{eq:5}
P(J_{ij})=\sqrt{\frac{N}{2\pi\sigma^{2}}}\exp-(\frac{NJ_{ij}^{2}}{2J^{2}}),
\end{eqnarray}
where $\frac{J}{\sqrt{N}}$ is the width of the distribution. The
replica trick is then implemented to get the free energy at the
large $N$ limit.

The case $r=1$, namely the random  exchange version of the
Blume-Capel model,  was first introduced and discussed by Ghatak
and Sherrington  (GS) \cite{Ghatak} who used symmetric replica to
obtain the relevant phase diagram. The GS solution seems to
display inverse freezing even for the $r=1$ case, but more
detailed analysis by da Costa et. al. \cite{Costa} revealed that
the glass order-parameter takes nonzero values (with a variety of
stability features) in the area below the GS transition line, and
the temperature dependence is monotonic. Recently, the full
replica symmetry breaking analysis has been implemented for the GS
model \cite{crisanti}, and the results admit no inverse glass
transition. Here we present a replica symmetric analysis of the
same hamiltonian where the interacting states are highly
degenerate, i.e., $r>1$. Following \cite{Costa}, we obtain the
phase transition and the spinodal lines, and the results support,
again, both first and second order inverse glass transition.

The replica technique \cite{Edwards} relies on the identity
$\overline{ln[Z]}=lim_{n \rightarrow 0}\frac{1}{n}(\overline{Z^n}
-1)$, where $Z$ is the partition function of the system  and $Z^n$
is interpreted as the partition function of an n-fold replicated
system $S_i \rightarrow S_{ia}, a=1...n$. The average free energy
may be computed using $\beta f =-lim_{n\rightarrow
0}\frac{1}{n}(\overline{Z^n}-1)$. The disorder average is taken
for $Z^n$ using the  Gaussian distribution (\ref{eq:5}) and gives:
\begin{eqnarray}
\label{eq:6} \overline {Z^{n}}= Tr \exp \left [\frac{\beta^2
J^2}{N} \sum_{a>b} (\sum_iS_{ia} S_{ib})^2+ \right. \nonumber\\
\left. \frac{\beta^2 J^2}{2N}\sum_{a}(\sum_i S_{ia}^2)^2-\beta D
\sum_{i} S_{ia} ^2  \right]
\end{eqnarray}
where $a,b=1...n$ denotes the replica. Implementing the
Hubbard-Stratanovitch identity yields the free energy per spin:
\begin{eqnarray} \label{ff}
\label{eq:7} -\beta \frac{F}{N}=-\beta^2J^2\sum_{a>b}q_{ab}^2\
-\frac{\beta^2J^2}{2}\sum_{a}q_{aa}^2+lnTr e^{ \hat{L}}
\end{eqnarray}
where
\begin{eqnarray}
\label{eq:8}
\hat{L}=2\beta^2J^2\sum_{a>b}q_{ab}S_aS_b+\beta^2J^2\sum_{a}q_{aa}S_{a}^2-\beta
D\sum_{a}S_{a}^2 .\nonumber\\
\end{eqnarray}
$q_{aa}$ and $q_{ab}$, the diagonal and the off diagonal entries
of the "order parameter matrix", are given self-consistently by
the saddle-point condition:
\begin{eqnarray}
\label{eq:9} q_{ab}&=&\langle{S_a S_b}\rangle\nonumber\\
 q_{aa}&=&\langle{S_a ^2}\rangle
\end{eqnarray}
where  $\langle ... \rangle$ stands for thermal average. In order
to solve this model it is necessary to make assumptions on the
order parameter matrix elements  $q_{ab}$, and the simplest
ansatz, is symmetry with respect to permutations of any pair of
the replicas: $q_{ab}=q, \ \  \forall a \neq b$, $q_{aa}=p, \ \
\forall a$. Using this \emph{replica symmetric} assumption one
obtains
\begin{eqnarray}\label{fff}
\label{eq:10} -\beta f&=&\frac{\beta^2 J^2}{2}(q^2-p^2)+
 \frac{1}{\sqrt{2\pi}}\int_{-\infty}^{\infty} dz \exp
(-\frac {z^2}{2})\cdot\nonumber\\ &ln&[1+2 \ r \ e^{\gamma}\cosh
(\beta J\sqrt{2q}z)
\end{eqnarray}
with
\begin{eqnarray}
\label{eq:11} \gamma=\beta^2 J^2 (p-q)-\beta D
\end{eqnarray}

\begin{figure}
  \includegraphics[width=7.7cm]{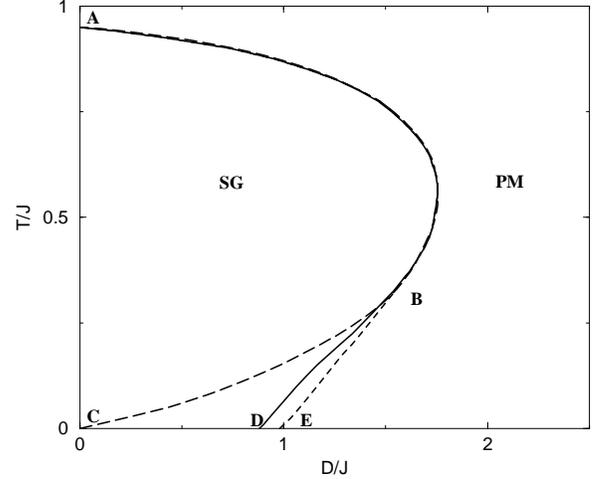}\\
  \caption{Phase diagram and the spinodal lines for the disordered model
  in the  D-T plane for a constant
  interaction J for $r=6$.}\label{fig5}
\end{figure}

Extremizing  the  free energy with respect to q and p one gets by
the following coupled equations:
\begin{eqnarray} \label{qqq}
\label{eq:12} q=\int_{-\infty}^{\infty}\frac{dz \ exp(-\frac
{z^2}{2})}{\sqrt{2\pi}}
 \frac {4 r ^2 e^{2\gamma}\sinh ^2(\beta J
\sqrt{2q}z)}{[1+2 r e^{\gamma}\cosh (\beta J \sqrt{2q}z)]^2}
\end{eqnarray}
\begin{eqnarray} \label{ppp}
\label{eq:13} p=\int_{-\infty}^{\infty} \frac{dz \ exp(-\frac
{z^2}{2})}{\sqrt{2\pi}}\frac {2 r e^{\gamma} \cosh(\beta J
\sqrt{2q}z)}{1+2 r e^{\gamma}\cosh (\beta J \sqrt{2q}z)}
\end{eqnarray}
The  coupled equations (\ref{qqq}) and (\ref{ppp}) are numerically
solved (with the possibility of multiple solutions if more than
one stable state exists), and the location of the first order
transition line is then determined by comparison of the free
energy   values (plugging $q$ and $p$ into (\ref{fff})).  The
resulting phase diagram is  shown in Fig. (\ref{fig5}) for the
case $r=6$, and displays all the essential features that exist in
the ordered model, including a tricritical point and spinodal
lines.

To conclude, the basic theoretical insight of Blume and Capel, to
have a spin model with low energy non-interacting (zero) state,
may yield an inverse melting transition once the model is enriched
with an entropic advantage of the interacting phase. It should be
emphasized that the  higher entropy associated with the
interaction do not unavoidably entail inverse melting; this
property may be "buried" below energetic and other constraints
that dominate the system, yet it may change the phase diagram
predicted by the naive assumption that higher energy implies
higher entropy.

The authors wish to acknowledge Prof. Y. Rabin for most helpful
discussions of the subject.

\end{document}